\documentclass[a4paper,11pt]{article}
\usepackage{pos}

\usepackage[normalem]{ulem}

\let\OLDthebibliography\thebibliography
\renewcommand\thebibliography[1]{
  \OLDthebibliography{#1}
  \setlength{\parskip}{0pt}
  \setlength{\itemsep}{0pt plus 0.3ex}
}

\usepackage{xcolor}
\usepackage{xparse}

\ExplSyntaxOn

\keys_define:nn { miguel/label }
 {
  label   .tl_set:N = \l_miguel_label_tl,
  unknown .code:n   = \clist_put_right:Nx \l_miguel_label_clist
                       { \l_keys_key_tl = \exp_not:n { #1 } }
 }
\clist_new:N \l_miguel_label_clist
\box_new:N \l_miguel_label_box
\box_new:N \l_miguel_label_image_box

\NewDocumentCommand{\xincludegraphics}{O{}m}
 {
  \tl_clear:N \l_miguel_label_tl
  \clist_clear:N \l_miguel_label_clist
  \keys_set:nn { miguel/label } { #1 }
  \tl_if_empty:NTF \l_miguel_label_tl
   {
    \miguel_includegraphics:Vn \l_miguel_label_clist { #2 }
   }
   {
    \hbox_set:Nn \l_miguel_label_image_box
     {
      \miguel_includegraphics:Vn \l_miguel_label_clist { #2 }
     }
    \hbox_set:Nn \l_miguel_label_box
     {
      \skip_horizontal:n { 3pt }
      \fcolorbox{black}{white}{\footnotesize \tl_use:N \l_miguel_label_tl}
     }
    \leavevmode
    \box_use:N \l_miguel_label_image_box
    \skip_horizontal:n { -\box_wd:N \l_miguel_label_image_box }
    \hbox_overlap_right:n
     {
      \box_move_up:nn
       {
        \box_ht:N \l_miguel_label_image_box - 
        \box_ht:N \l_miguel_label_box - 3pt
       }
       { \box_use_drop:N \l_miguel_label_box }
     }
    \skip_horizontal:n { \box_wd:N \l_miguel_label_image_box }
   }
 }

\cs_new_protected:Nn \miguel_includegraphics:nn
 {
  \includegraphics[#1]{#2}
 }
\cs_generate_variant:Nn \miguel_includegraphics:nn { V }

\ExplSyntaxOff

\title{New insights from old cosmic rays:\\ A novel analysis of archival KASCADE data}
 \ShortTitle{A novel analysis of archival KASCADE data}

\author*[a]{D.~Kostunin}
\author[b,c]{I.~Plokhikh}
\author[d]{M.~Ahlers}
\author[e]{V.~Tokareva}
\author[e]{V.~Lenok}
\author[f]{P.~Bezyazeekov}
\author[g]{S.~Golovachev}
\author[g]{V.~Sotnikov}
\author[b,c]{R.~Mullyadzhanov}
\author[h]{E.~Sotnikova}

\affiliation[a]{DESY, 15738 Zeuthen, Germany}
\affiliation[b]{Novosibirsk State University, 630090 Novosibirsk, Russia}
\affiliation[c]{Institute of Thermophysics SB RAS, 630090 Novosibirsk, Russia}
\affiliation[d]{Niels Bohr Institute, University of Copenhagen, DK-2100 Copenhagen, Denmark}
\affiliation[e]{Karlsruhe Institute of Technology, Institute for Astroparticle Physics, 76021 Karlsruhe, Germany}
\affiliation[f]{Applied Physics Institute, Irkutsk State University, 664020 Irkutsk, Russia}
\affiliation[g]{JetBrains Research, 194100 St. Petersburg, Russia}
\affiliation[h]{Sobolev Institute of Mathematics, 630090 Novosibirsk, Russia}

\emailAdd{astroparticle@jetbrains.com}

\abstract{
Cosmic ray data collected by the KASCADE air shower experiment are competitive in terms of quality and statistics with those of modern observatories. We present a novel mass composition analysis based on archival data acquired from 1998 to 2013 provided by the KASCADE Cosmic ray Data Center (KCDC). The analysis is based on modern machine learning techniques trained on simulation data provided by KCDC. We present spectra for individual groups of primary nuclei, the results of a search for anisotropies in the event arrival directions taking mass composition into account, and search for gamma-ray candidates in the PeV energy domain.
}

\FullConference{37$^{\rm{th}}$ International Cosmic Ray Conference (ICRC 2021)\\
		July 12th -- 23rd, 2021\\
		Online -- Berlin, Germany}

\begin{document}
\maketitle

\section{Introduction}\label{sec1}

Our knowledge of cosmic rays (CRs) remains sketchy even one century after their discovery, see {\it e.g.}~\cite{Gabici:2019jvz,AlvesBatista:2019tlv}. In particular, the transition between the dominance of Galactic and extragalactic sources that is expected to occur somewhere between the CR ``knee'' and ``ankle'' is highly uncertain. There are various approaches to unravel this mystery. Besides the indirect observation via hadronic $\gamma$-ray and neutrino emission produced in CR interactions, precision measurements of CR spectra can provide valuable information, especially when their masses and charges are measured with high resolution.

The charge and mass reconstruction of high-energy CRs above the knee, which are observable via extended air-showers, is a special problem since the correlation of primary charge and mass with distribution of secondary particles is weak. In addition, the reconstruction depends strongly on hadronic interaction models introducing additional systematic uncertainties. This is the main reason for skepticism regarding prior results obtained using simplified methods and interpreted with, by now, outdated models, despite the huge amount of analyzed data.

In these proceedings, we present the results of a novel analysis of archival data collected by KASCADE~\cite{KASCADE:2003swk} and provided by the KCDC service~\cite{Haungs:2018xpw} based on the latest hadronic interaction models and machine-learning techniques. The reconstruction of the primary charge allows us to obtain CR spectra of individual mass groups and to study CR anisotropies in terms of particle rigidity. We also report the first steps towards the search for photons in KASCADE data.

We will start in section~\ref{sec2} with a description of our methods based on a random forest, which is an ensemble machine learning method, using sets of decision trees on various sub-samples of training data to improve accuracy compared with single decision tree. We present our results on mass composition in Section~\ref{sec3} followed by an analysis of rigidity-dependent large-scale anisotropies in Section~\ref{sec4}. We present our results on the photon fraction in Section~\ref{sec5} before we conclude in Section~\ref{sec6}.

\section{Methods and Data}\label{sec2}

We used KASCADE preselection data sets\footnote{\url{https://kcdc.iap.kit.edu/datashop/fulldata/}}, which contain the following reconstructed air shower properties that we use for the training of a classifier of primary mass: energy $E$; shower core coordinates $(x,y)$; arrival direction $(\theta, \phi)$; muon and electron numbers $\log_{10} N_\mu$, $\log_{10} N_e$; and shower age $s$. The KDCD service provides CORSIKA~\cite{Heck:1998vt} simulations with events generated for five individual mass groups: \textit{H}, \textit{He}, \textit{C}, \textit{Si}, \textit{Fe}. These simulations provide the same properties as in real data reconstructed using the actual detector response. We have trained a classifier to return one of the five mass groups based on three modern hadronic interaction models: QGSJet-II.04~\cite{Ostapchenko:2010vb}, EPOS-LHC~\cite{Pierog:2013ria} and Sibyll 2.3c~\cite{Riehn:2015aqb}.

\begin{figure}[t]
\centering
\includegraphics[width=0.32\linewidth]{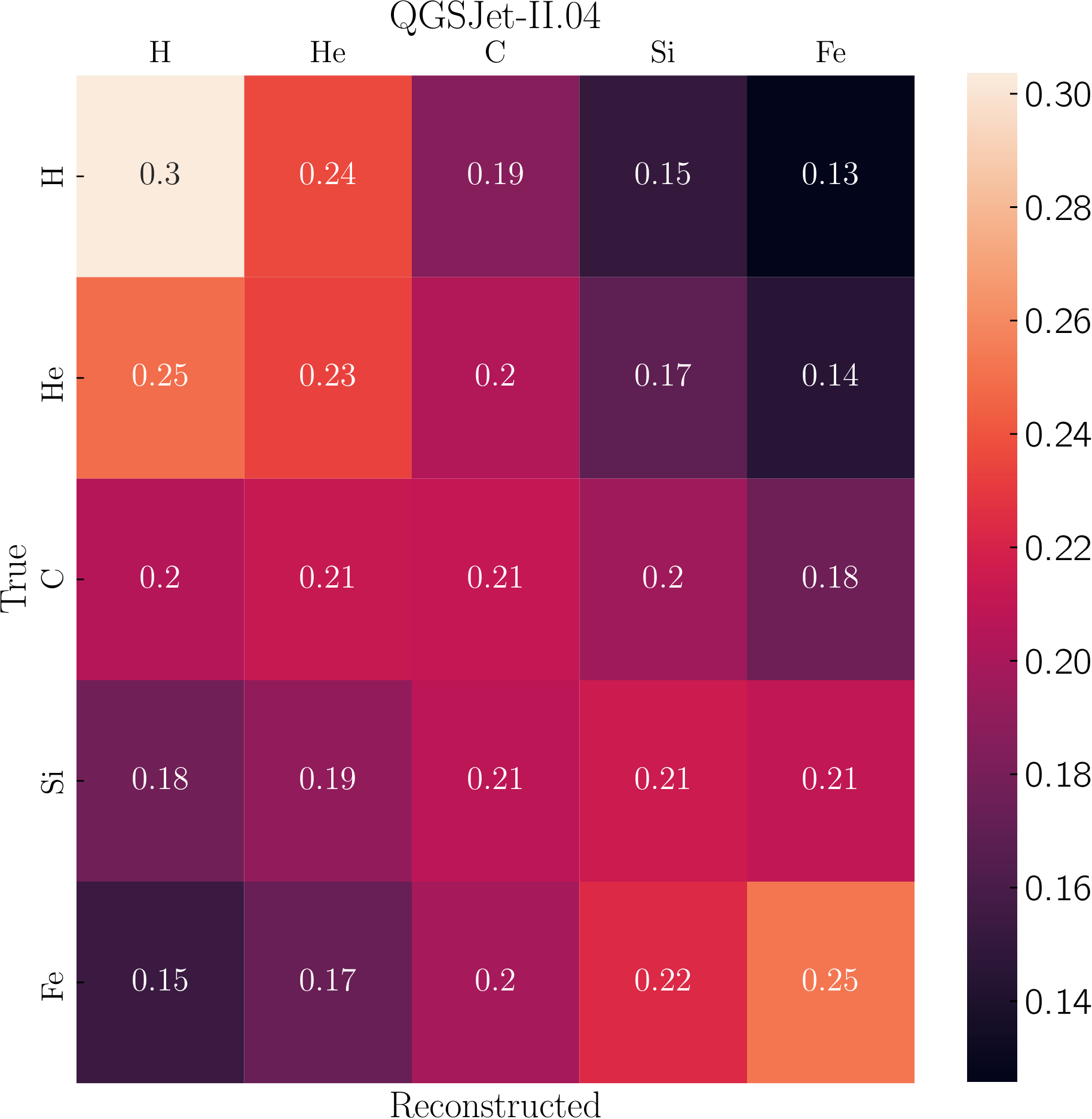}\hfill
\includegraphics[width=0.32\linewidth]{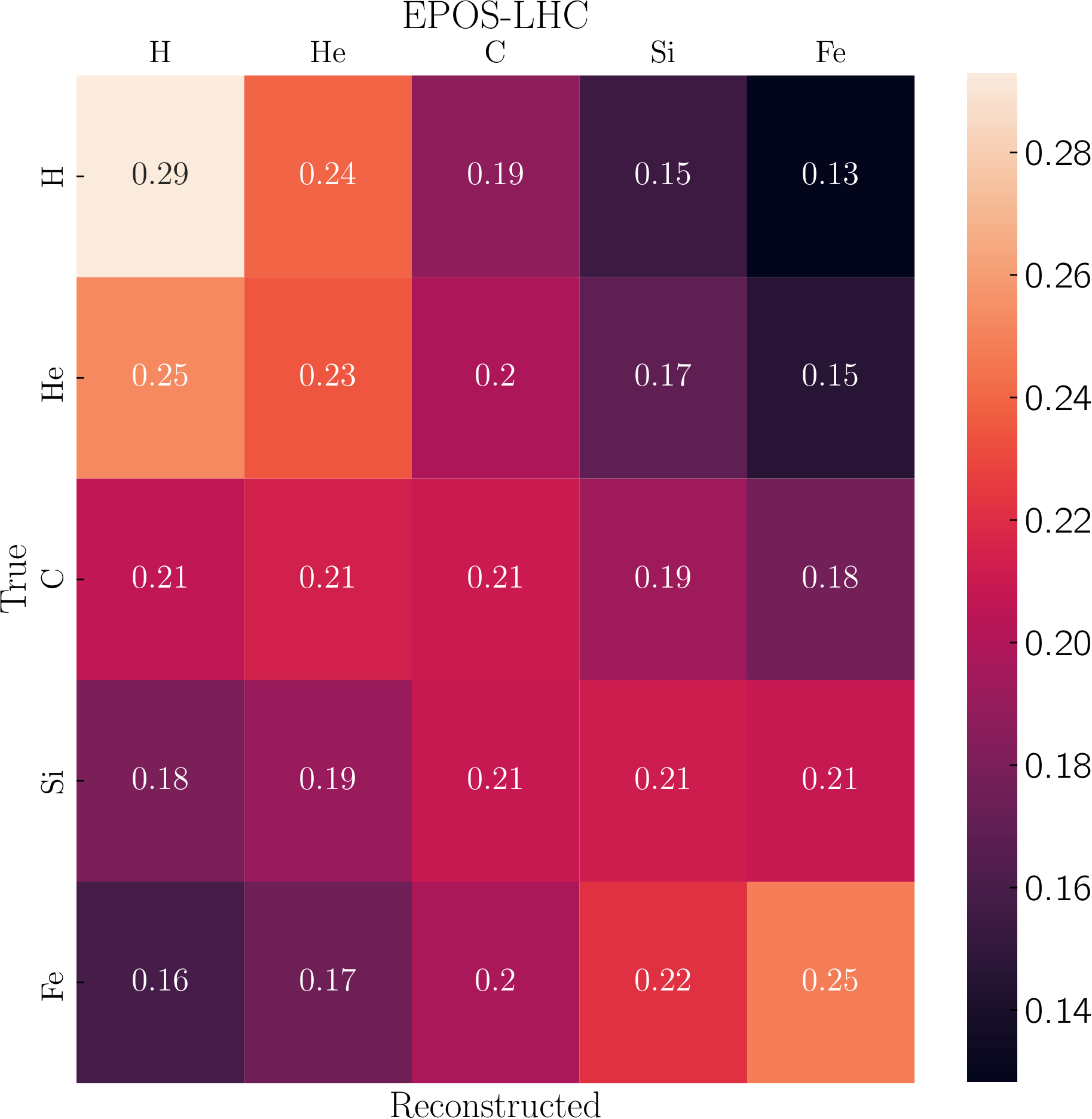}\hfill
\includegraphics[width=0.32\linewidth]{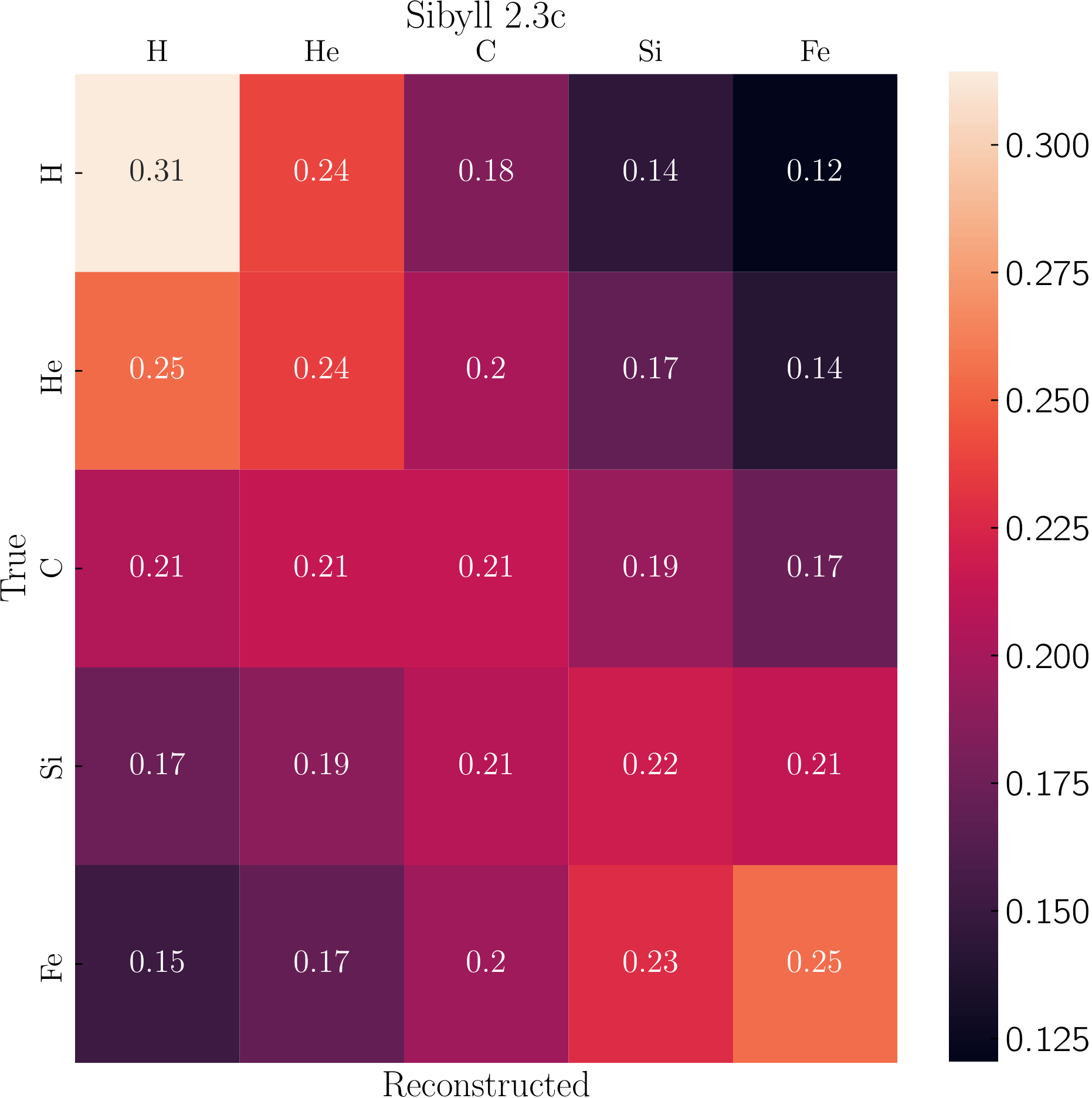}
\caption[]{Confusion matrix for our classifier trained using three different hadronic interaction models, QGSJet-II.04~\cite{Ostapchenko:2010vb}, EPOS-LHC~\cite{Pierog:2013ria} and Sibyll 2.3c~\cite{Riehn:2015aqb}.}
\label{fig:cm}
\end{figure}

The confusion matrix of the classifier is shown in Fig.~\ref{fig:cm}.
One can see that the matrix has a diagonal structure; however, non-diagonal elements are heavily contaminated, which indicates systematic uncertainties. 
It is worth noting, that these matrices obtained for the simulation data before application of any quality cuts. 
In our work we used the quality cuts defined by KASCADE collaboration~\cite{KASCADE:2005ynk}, suggesting the following: $x^2+y^2<91$\,m, $\log_{10} N_\mu\ge3.6$, $\log_{10} N_e\ge4.8$, $0.2<s<2.1$, $\theta<18^\circ$. 
Simulation study does not show any degradation of the classifier performance, but reconstructed spectra indicate irregularities, which might point to discrepancies between simulations and data beyond official quality cuts.

\begin{figure}[ph!]
\centering
\xincludegraphics[width=0.48\linewidth]{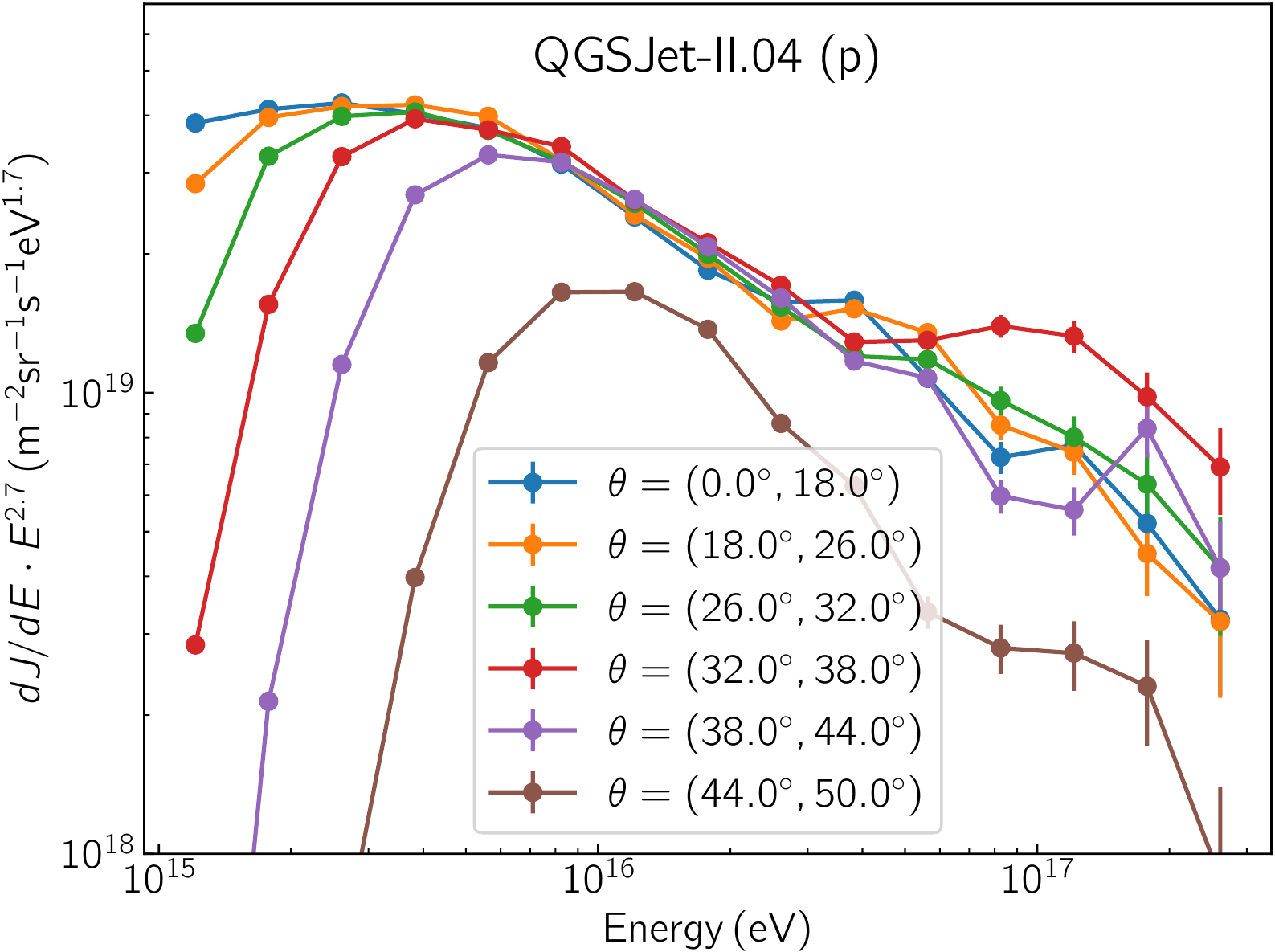}\hspace{0.5cm}
\includegraphics[width=0.48\linewidth]{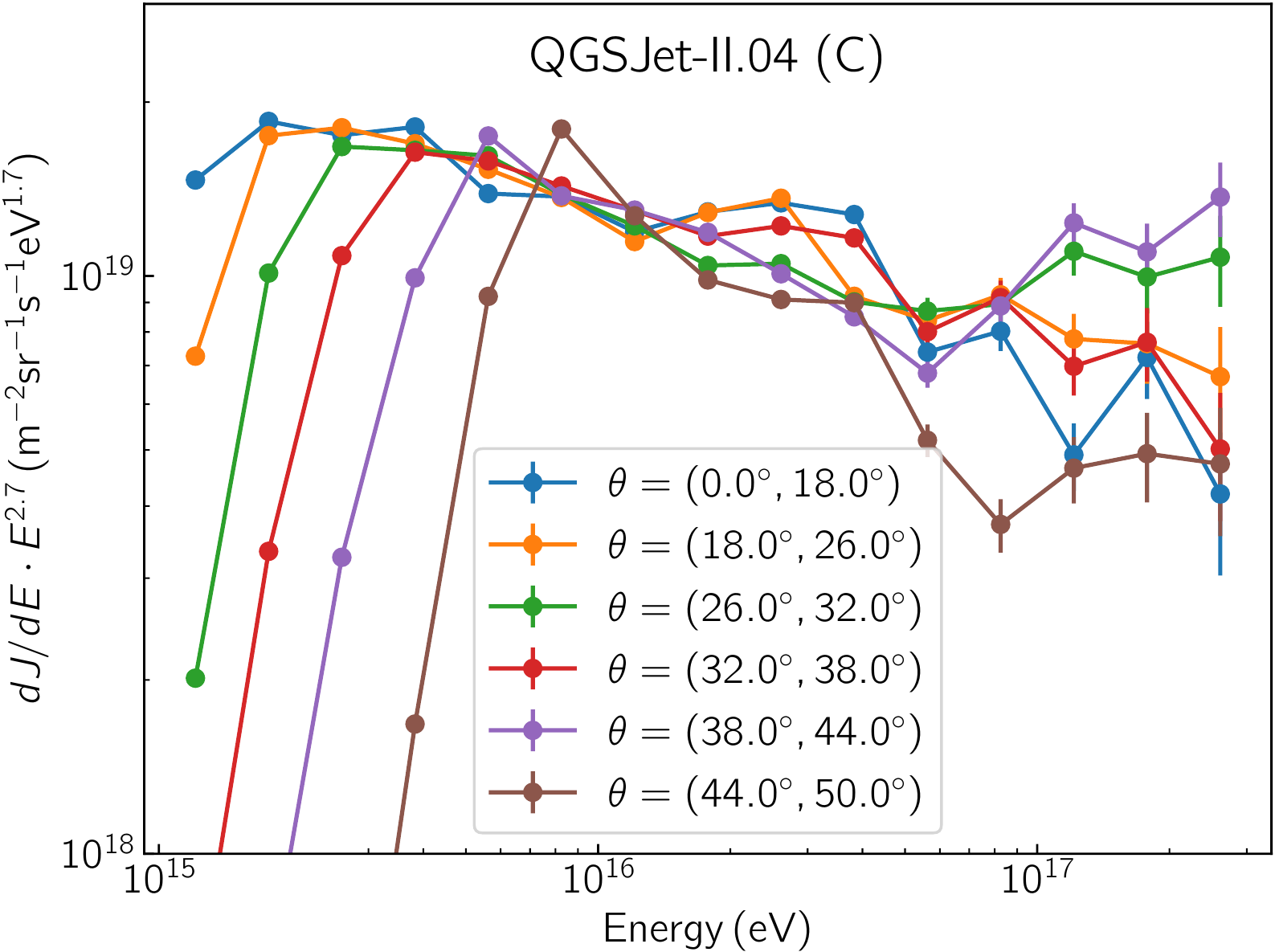}
\caption[]{Spectra of primary hydrogen (left) and carbon (right) mass groups reconstructed with QGSJet-II.04 using zenith angles beyond KASCADE quality cuts. 
The statistical uncertainties for heavier mass groups are too large to allow for a study of systematic effects. The zenith bands are selected in order to obtain equal exposure for each curve. The results indicate that the zenith angle cut might be accurately pushed to $\mathcal{O}(30^\circ)$, thereby increasing the exposure by a factor $\simeq 3$.}
\label{fig:zenith_comp}
\end{figure}

For the study of CR anisotropies it is interesting to go for larger zenith angles in order to cover a broader declination range. Figure~\ref{fig:zenith_comp} shows the spectra of primary hydrogen (left panel) and carbon (right panel) mass groups reconstructed with QGSJet-II.04 for zenith angles beyond $18^\circ$. The spectra suggest that the zenith angle cut can be extended $\mathcal{O}(30^\circ)$. For this reason, we present spectra for a conservative zenith angle cut of $\theta<18^\circ$, while we allow for a somewhat looser cut of $\theta<30^\circ$ in our anisotropy study. 

\section{Cosmic-Ray Mass Composition}\label{sec3}

\begin{figure}[ph!]
\centering
\includegraphics[width=0.48\linewidth]{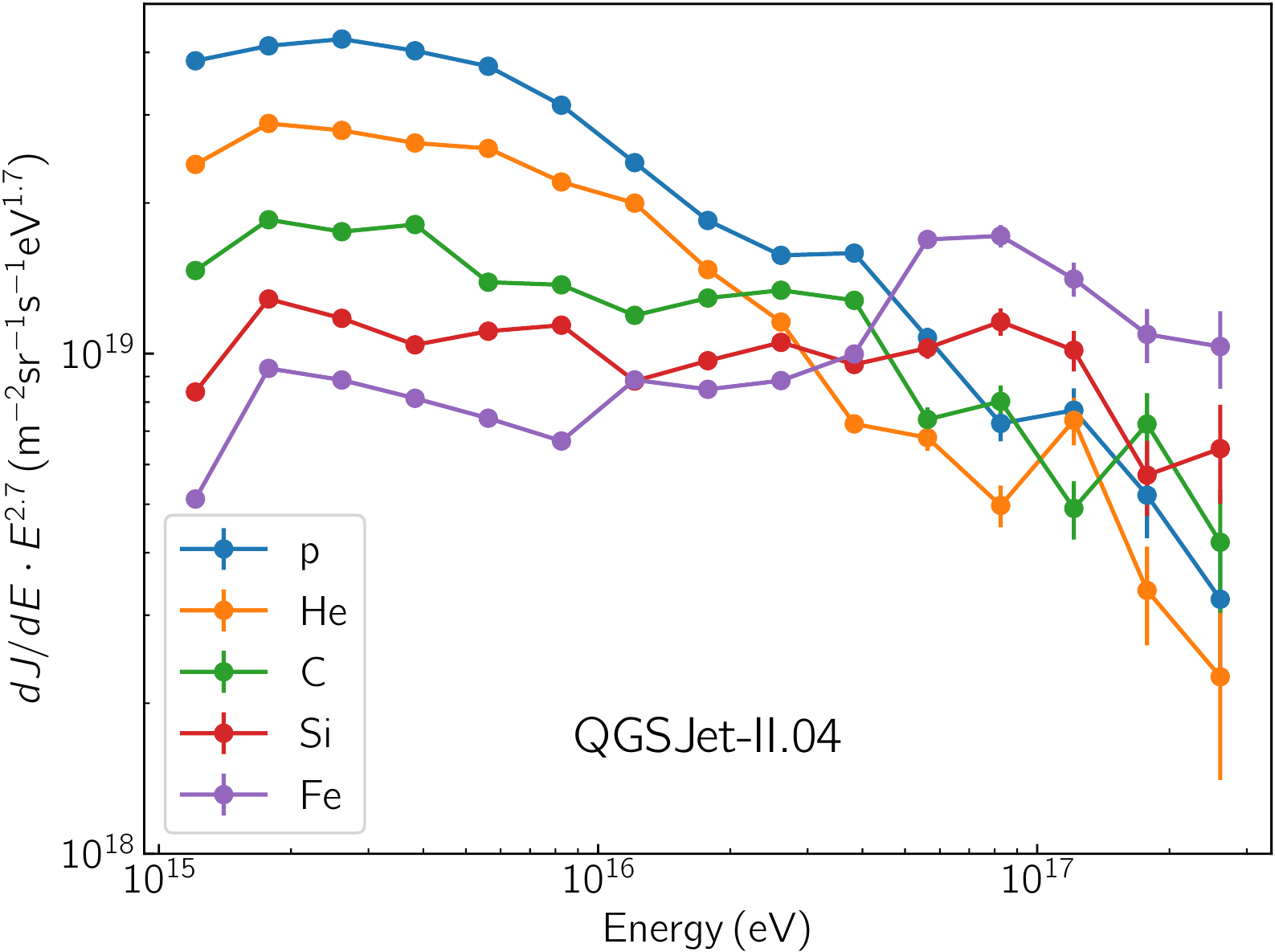}\hspace{0.5cm}
\includegraphics[width=0.48\linewidth]{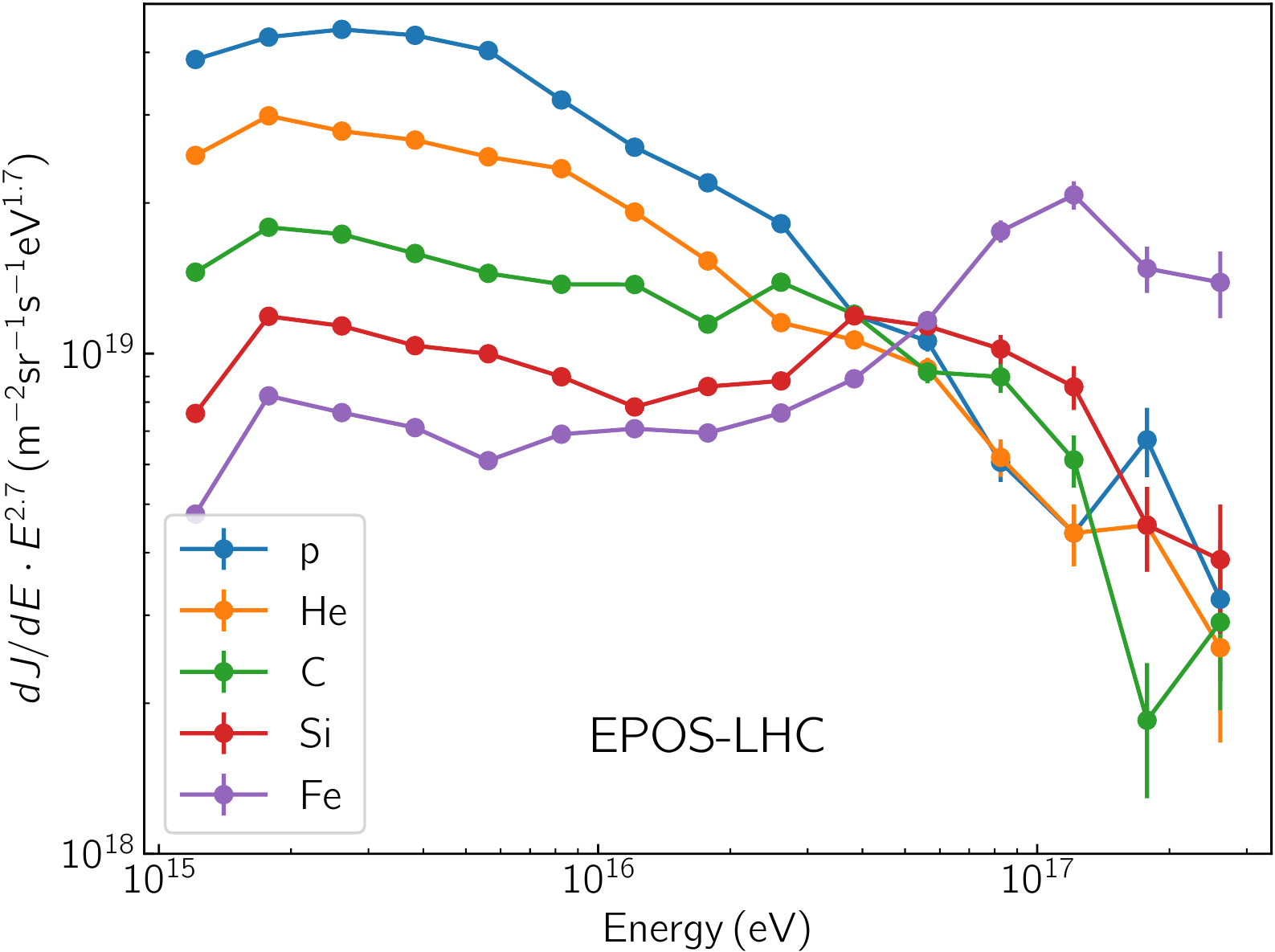}\\[0.3cm]
\includegraphics[width=0.48\linewidth]{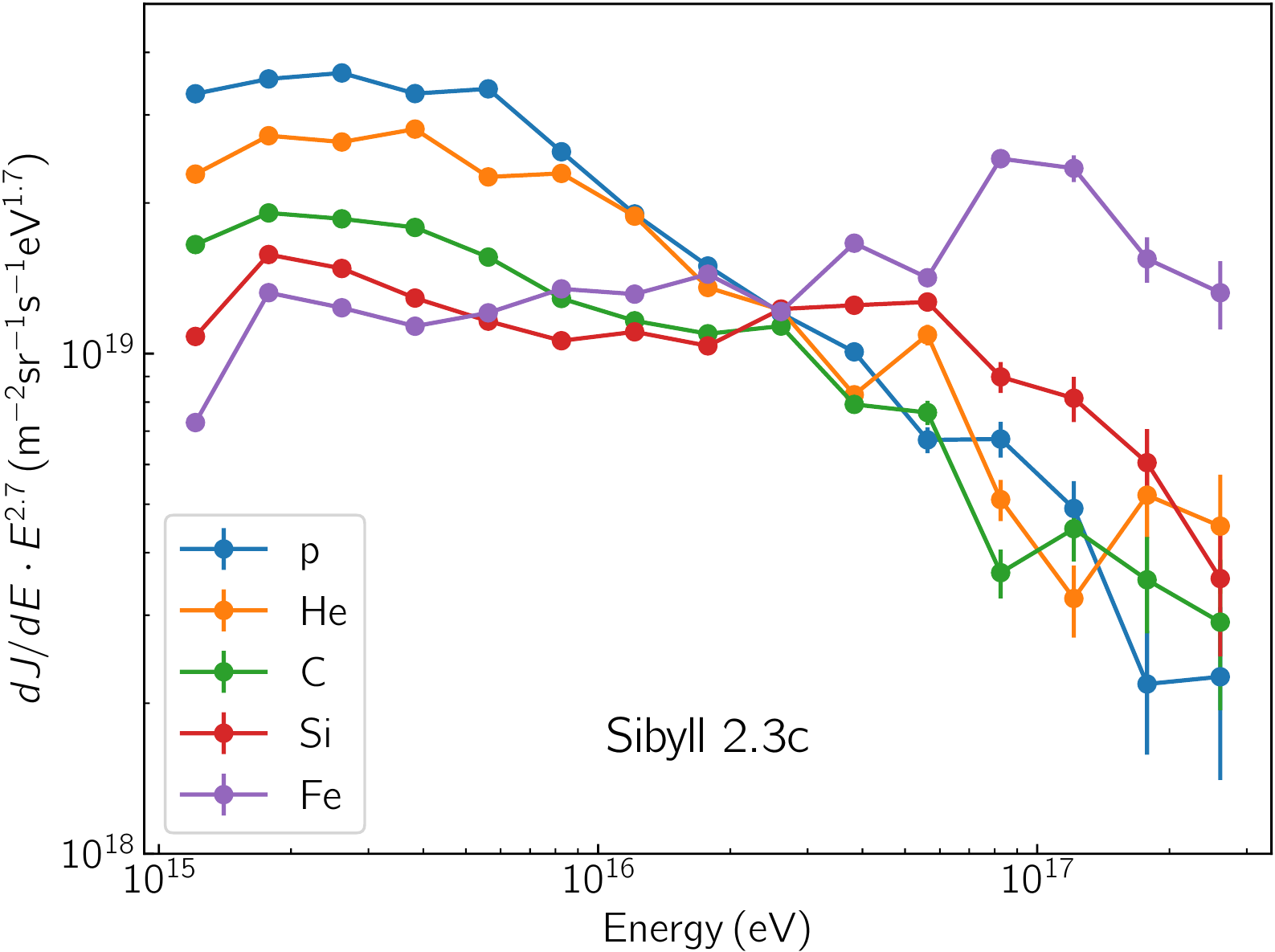}\hspace{0.5cm}
\includegraphics[width=0.48\linewidth]{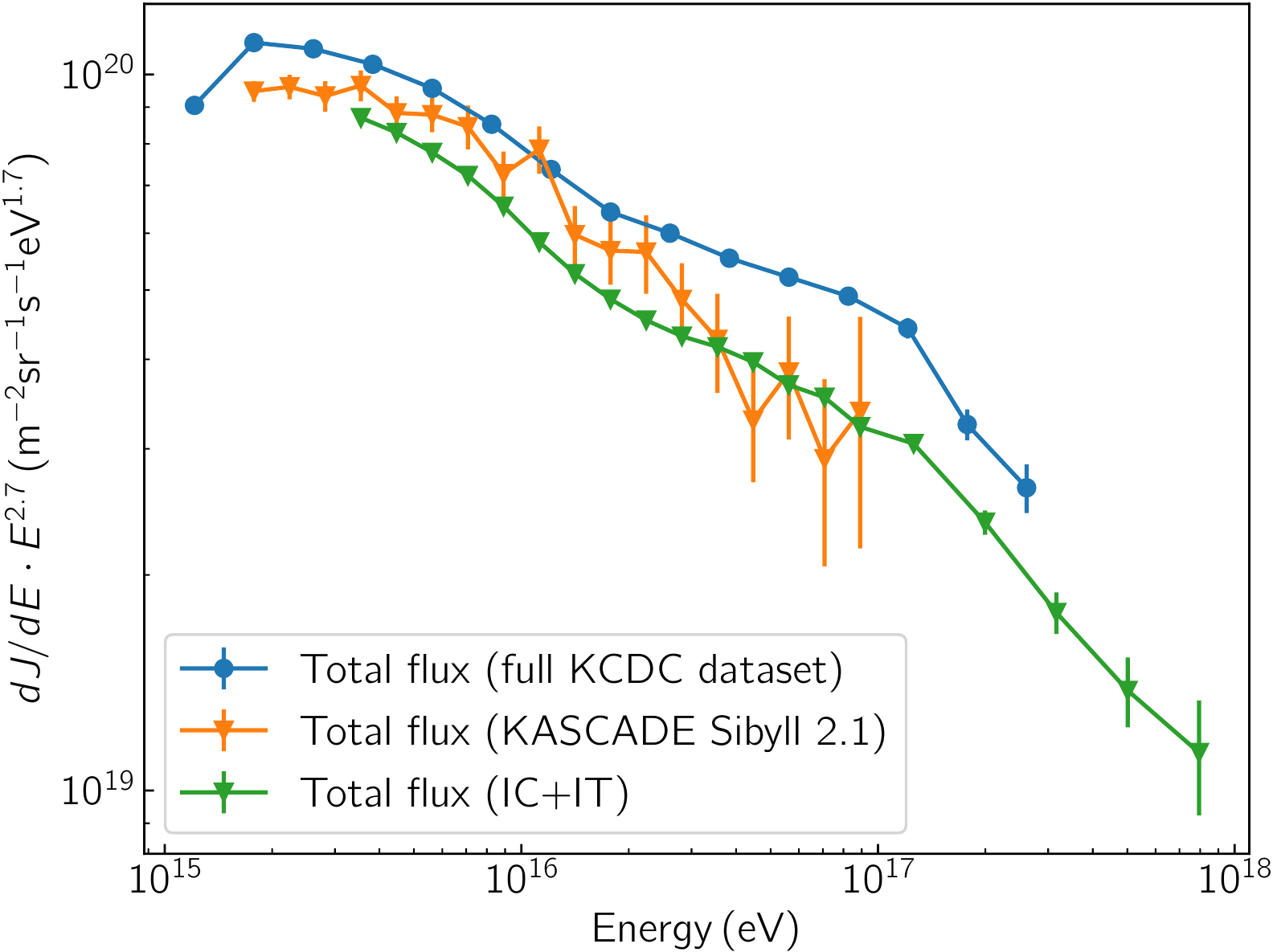}
\caption[]{Cosmic ray spectra for five individual mass groups and their sum reconstructed from the full KCDC data (without spectral unfolding) using different hadronic interaction models. We compare our results to those derived by KASCADE~\cite{KASCADE:2005ynk} and IceCube/IceTop~\cite{IceCube:2019hmk}.}
\label{fig:mass_comp_ours}
\end{figure}

For the reconstruction of mass-dependent CR spectra we assume full efficiency of the detector for events that pass the high-quality cuts described above. The livetime of the detector is obtained from the fit of time differences between two consecutive events with exponential decay function, {\it i.e.}~assuming a Poisson process. The full dataset corresponds to a livetime of $T_\mathrm{live} \simeq 0.42\times10^9$\,s. Thus, the total exposure has the form $\mathcal{E} \simeq \pi\sin^2\theta_{\rm max} \times S \times T_\mathrm{live}$, where $\theta_{\rm max}=18^\circ$ is the maximum zenith angle and $S = \pi (91{\rm m})^2$ is the surface area of the detector after quality cuts. 
Figure~\ref{fig:mass_comp_ours} shows the CR spectra reconstructed using three different hadronic interaction models. We show results for individual mass groups and the total flux. These results are consistent with earlier findings that reconstructions based on Sibyll show a trend towards heavier mass compositions. 

\begin{figure}[t]
\centering
\includegraphics[width=0.45\linewidth]{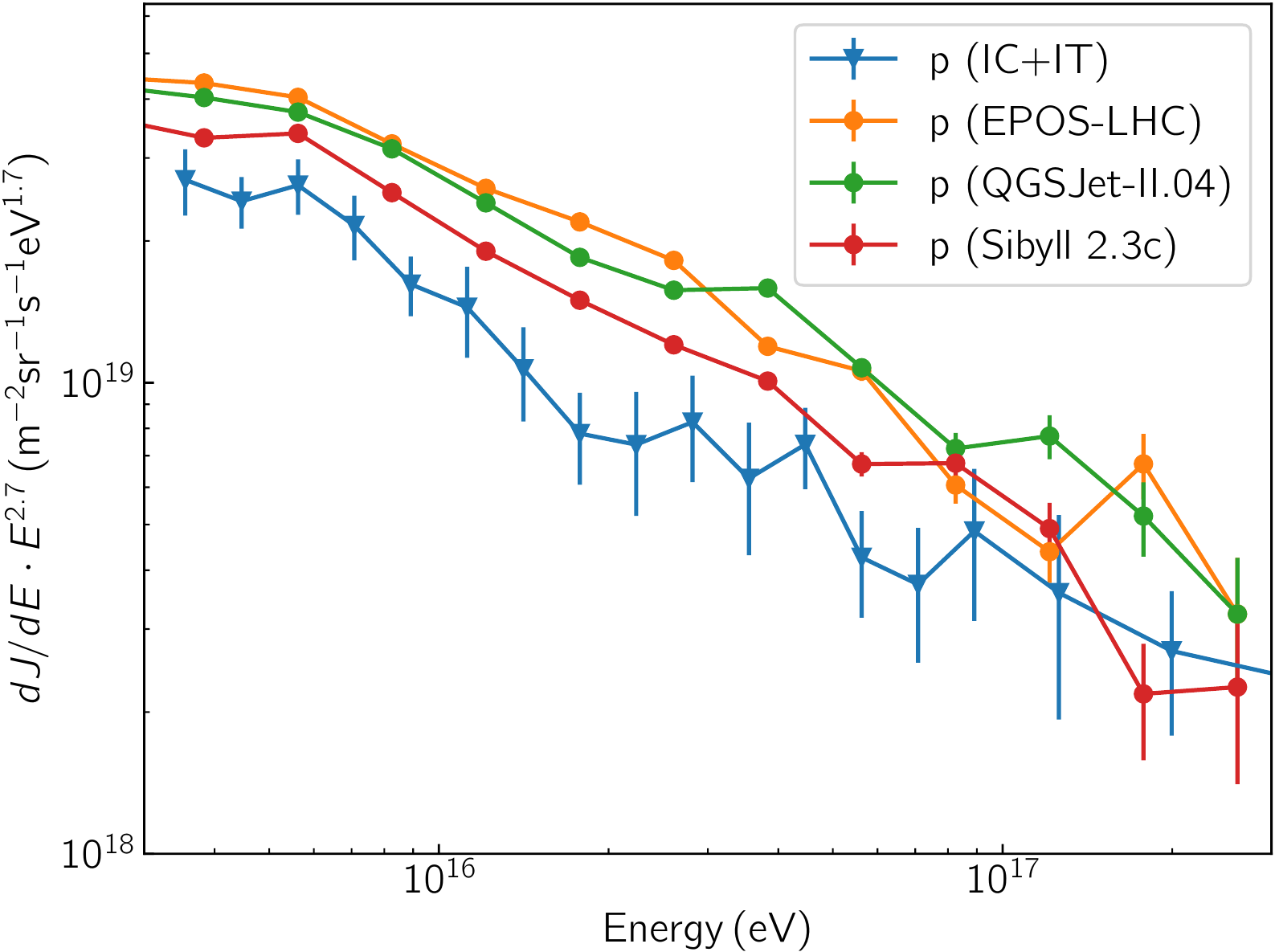}\hspace{0.5cm}
\includegraphics[width=0.45\linewidth]{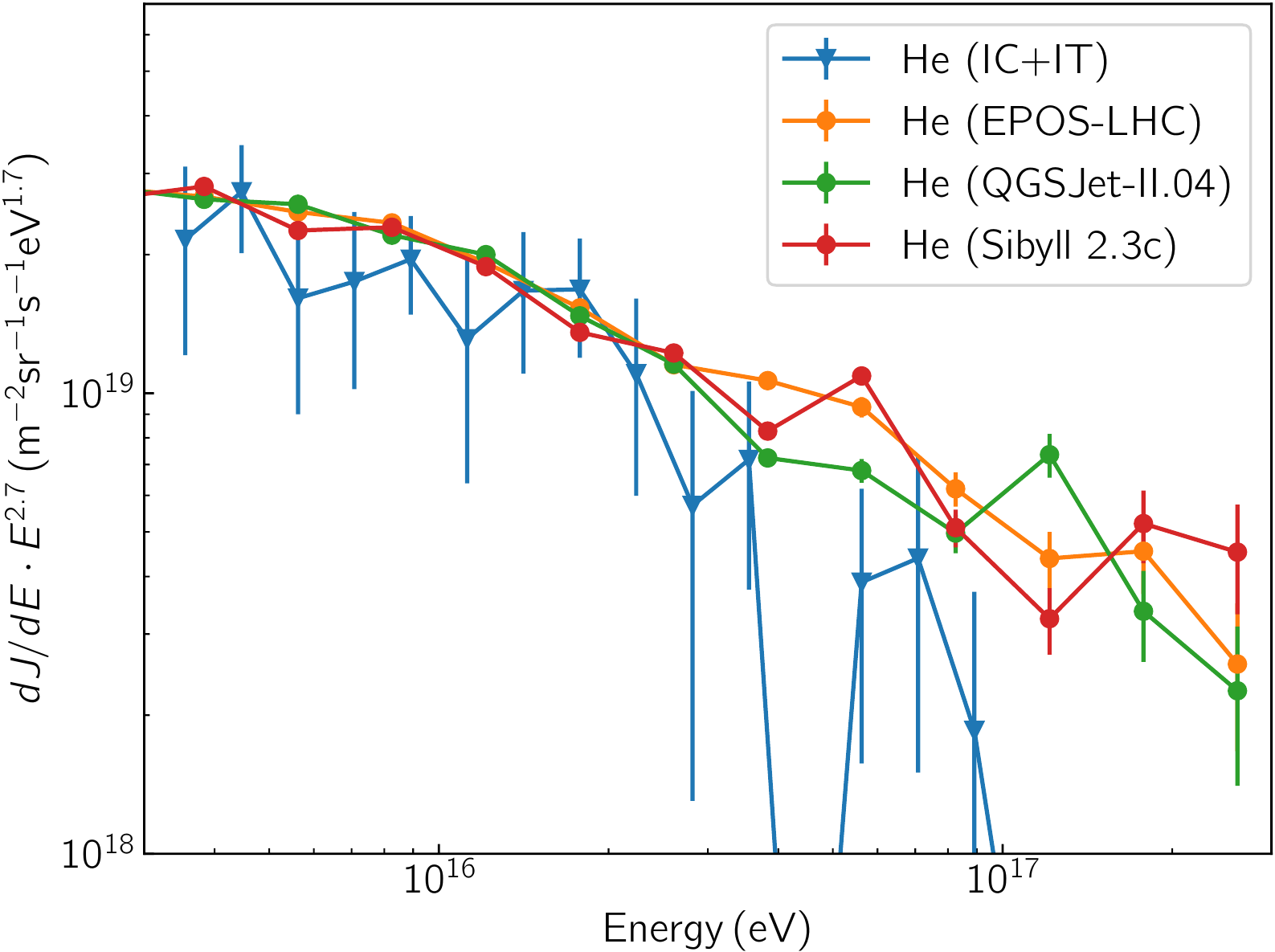}\\[0.3cm]
\includegraphics[width=0.45\linewidth]{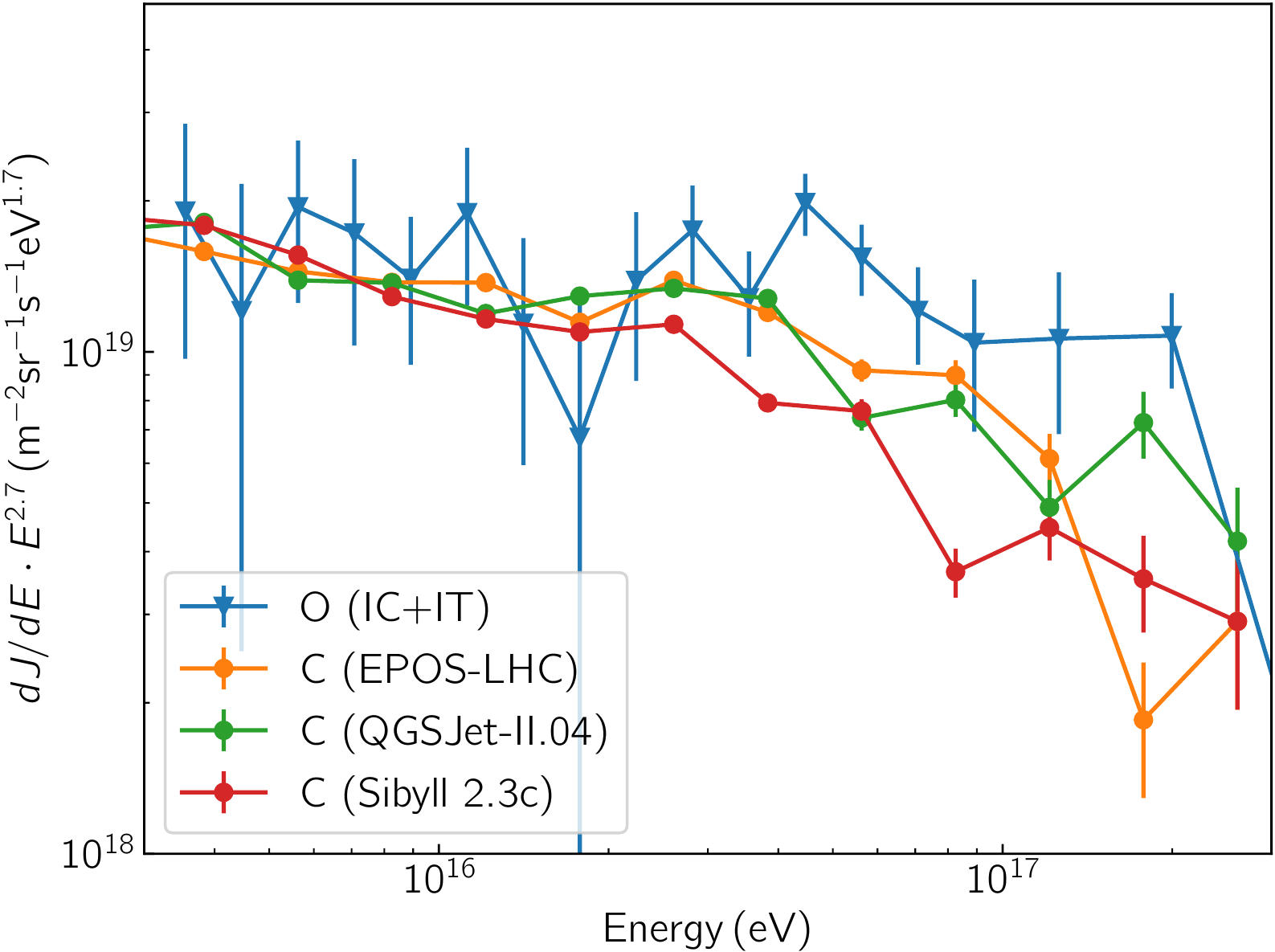}\hspace{0.5cm}
\includegraphics[width=0.45\linewidth]{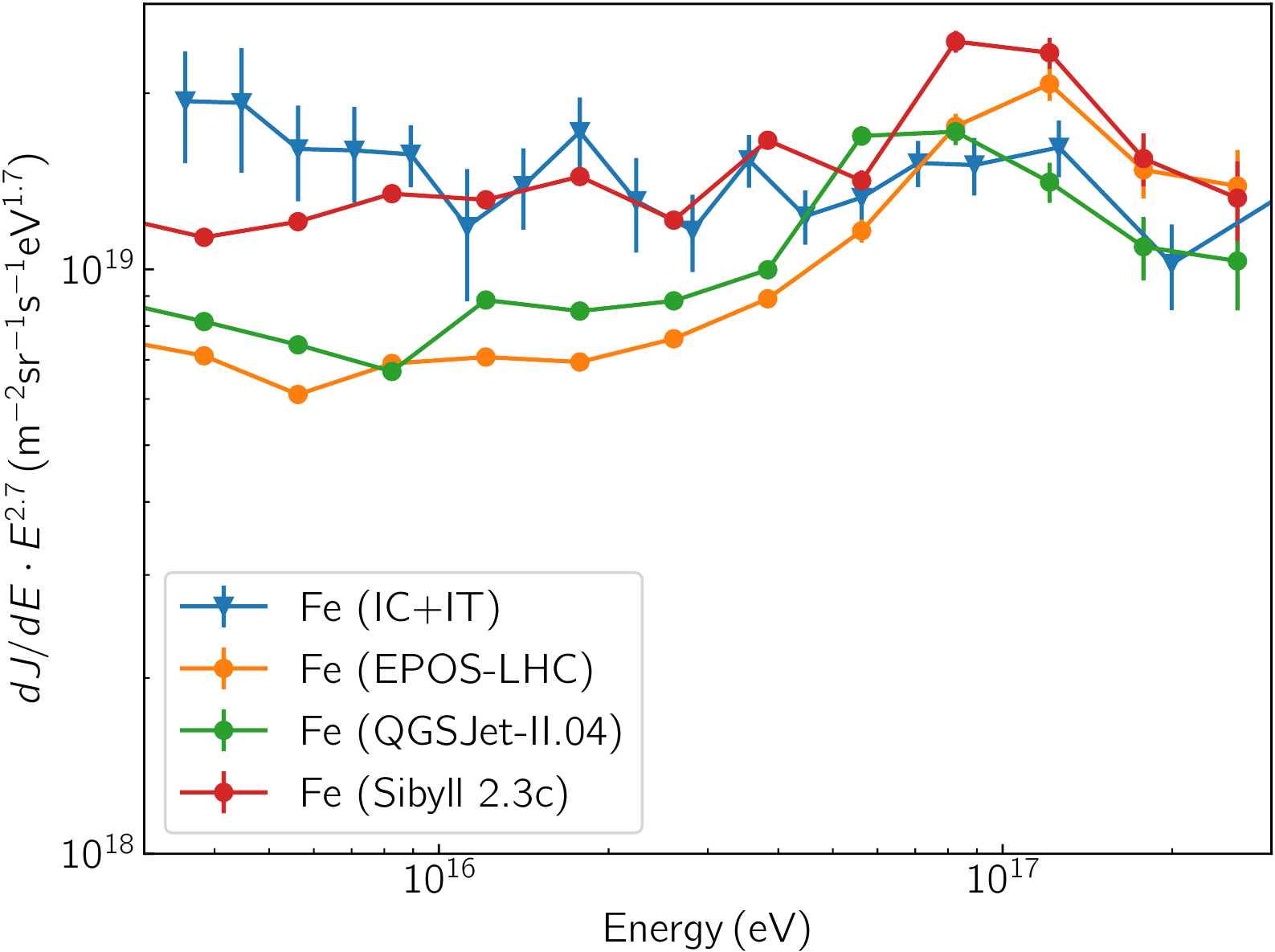}
\caption[]{Comparison of spectra for four mass groups as provided by IceCube/IceTop~\cite{IceCube:2019hmk} using Sibyll~2.1. It is important to point out that our analysis reconstructs the spectra of five different mass groups instead of only four used in the case IceCube/IceTop. The data corresponding to the spectrum of the \textit{Si} group (not presented here) should be redistributed between the remaining four groups according to the classifier confusion matrix described above. This lack of events is clearly visible for the \textit{C}/\textit{O} and \textit{Fe} groups, where contamination from \textit{Si} is more significant than for the lighter components.}
\label{fig:ic_comp}
\end{figure}

Figure~\ref{fig:ic_comp} shows a comparison with recent results by IceCube/IceTop~\cite{IceCube:2019hmk} based on Sibyll~2.1. We have chosen this particular study for comparison, because it provides results on four mass groups \textit{H}, \textit{He}, \textit{O}, and \textit{Fe}, which allows us to make a more accurate and direct comparison of our results on \textit{H}, \textit{He}, \textit{C}, and \textit{Fe}. The spectra of light components, \textit{H} \& \textit{He}, are consistent between experiments within uncertainties. The differences that we see between \textit{C} and \textit{O} and between the individual spectra for \textit{Fe} might be related to a ``contamination'' with \textit{Si}, that is not accounted for in the IceCube/IceTop analysis. A detailed study of this effect is out of scope of this work. Another important point, which should be kept in mind, is that we used default energy reconstructions from KCDC, which is not corrected for the mass composition, so our spectra are not completely unfolded.

\begin{table}[ph!]
\renewcommand{\arraystretch}{1.3}\renewcommand{\tabcolsep}{5pt}
\centering
\begin{tabular}{ccccccc}
\hline
$\mathcal{R}$ [${\rm V}$]&model&$N_{\rm tot}$&$A$ [$10^{-3}$]&$\alpha$ [${}^\circ$]&$p$-value&$A_{90}$~[$10^{-3}$]\\
\hline
$[10^{15.5},10^{16.0}]$&EPOS LHC&$897,294$&$10.1^{+5.5}_{-3.5}$&$251\pm28$&$0.10$&$17.1$\\
$> 10^{16.0}$&EPOS LHC&$79,140$&$19.6^{18.0}_{-8.4}$&$272\pm48$&$0.47$&$44.3$\\
\hline
$[10^{15.5},10^{16.0}]$&QGSJET-II.04&$874,416$&$14.3^{+5.4}_{-3.9}$&$278\pm20$&$0.01$&$21.1$\\
$> 10^{16.0}$&QGSJET-II.04&$74,665$&$18.7^{+18.5}_{-8.0}$&$234\pm51$&$0.52$&$44.3$\\
\hline
$[10^{15.5},10^{16.0}]$&Sibyll 2.3c&$753,824$&$7.7^{+5.9}_{-3.2}$&$261\pm40$&$0.33$&$15.6$\\
$> 10^{16.0}$&Sibyll 2.3c&$65,097$&$14.3^{+20.5}_{-5.1}$&$278\pm67$&$0.71$&$42.7$\\
\hline
\end{tabular}
\caption[]{Reconstructed dipole anisotropy using maximum-likelihood techniques discussed in Refs.~\cite{Ahlers:2018qsm,Ahlers:2019gdc}. Column 4 and 5 show the best-fit amplitude and phase of the sidereal dipole anisotropy with 68\% uncertainty range. The last column shows the 90\% C.L.~upper limit on the amplitude.}\label{tab1}
\end{table}

\begin{figure}[ph!]\centering
\includegraphics[width=0.8\linewidth,viewport=60 350 540 370,clip=true]{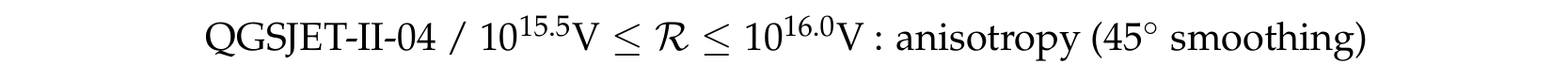}\\
\includegraphics[width=0.8\linewidth,viewport=5 200.5 595 350,clip=true]{./anisotropy/CR_bin1_45deg_qgs-4.pdf}\\
\includegraphics[width=0.8\linewidth,viewport=45 0 555 45,clip=true]{./anisotropy/CR_bin1_45deg_qgs-4.pdf}\\
\includegraphics[width=0.8\linewidth,viewport=60 350 540 370,clip=true]{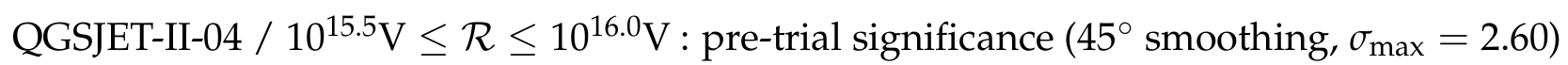}\\
\includegraphics[width=0.8\linewidth,viewport=5 200.5 595 350,clip=true]{./anisotropy/pretrial_bin1_45deg_qgs-4.pdf}\\
\includegraphics[width=0.8\linewidth,viewport=45 0 555 45,clip=true]{./anisotropy/pretrial_bin1_45deg_qgs-4.pdf}
\caption[]{Mollweide projections in equatorial coordinates of the reconstructed anisotropy (top) and pre-trial significance (bottom) for the rigidity bin $10^{15.5}<\mathcal{R}/{\rm V}<10^{16.0}$
based on QGSJET-II.04. We show the results for a top-hat smoothing radius of $45^\circ$. The grey-shaded area indicates the unobservable part of the celestial sphere. The dashed line indicates the projection of the Galactic Plane. The values of pre-trial significance are shown in units of standard deviations and indicated in red and blue colors for excesses and deficits, respectively. The location of maximum pre-trial significance is indicated by the symbol $\boldsymbol\times$. The anisotropy reconstruction is based on a maximum-likelihood method introduced in Ref.~\cite{Ahlers:2016njl}.}\label{fig1}
\end{figure}

\section{Rigidity-Dependent Anisotropy}\label{sec4}

The reconstruction of CR composition allows us -- for the first time -- to analyze the anisotropy of CR arrival direction in terms of rigidity~$\mathcal{R} = pc/Ze$. Table~\ref{tab1} shows the results of the sidereal dipole anisotropy using data with zenith angle $\theta\leq30^\circ$. We bin the data into two rigidity bins, $10^{15.5}{\rm V}<\mathcal{R} < 10^{16.0}{\rm V}$ and $10^{16}{\rm V}<\mathcal{R}$, based on the average charge of the five individual mass groups inferred by the composition analysis. The dipole analysis is based on a maximum-likelihood method following Refs.~\cite{Ahlers:2018qsm,Ahlers:2019gdc}. We do not find strong evidence for large-scale anisotropies and place 90\% C.L.~upper limits on the dipole amplitude (last column of Table~\ref{tab1}).

The most significant excess with a $p$-value of $0.01$ is found for the first rigidity bin using the composition based QGSJET-II.04. Figure~\ref{fig1} shows the corresponding relative intensity (top panel) and pre-trial significance (bottom panel) averaged over a radius of $45^\circ$. These all-sky results are based on a maximum-likelihood method introduced in Ref.~\cite{Ahlers:2016njl}. The smoothed relative intensity is consistent with the best-fit orientation of the dipole anisotropy. Note that the $45^\circ$ smoothing scale reduces the amplitude of the excess compared to the dipole fit shown in Table~\ref{tab1}. These results are consistent with previous anisotropy measurements; see {\it e.g.}~Ref.~\cite{Ahlers:2016rox}.

\begin{figure}[t]
\centering
\includegraphics[height=0.31\linewidth]{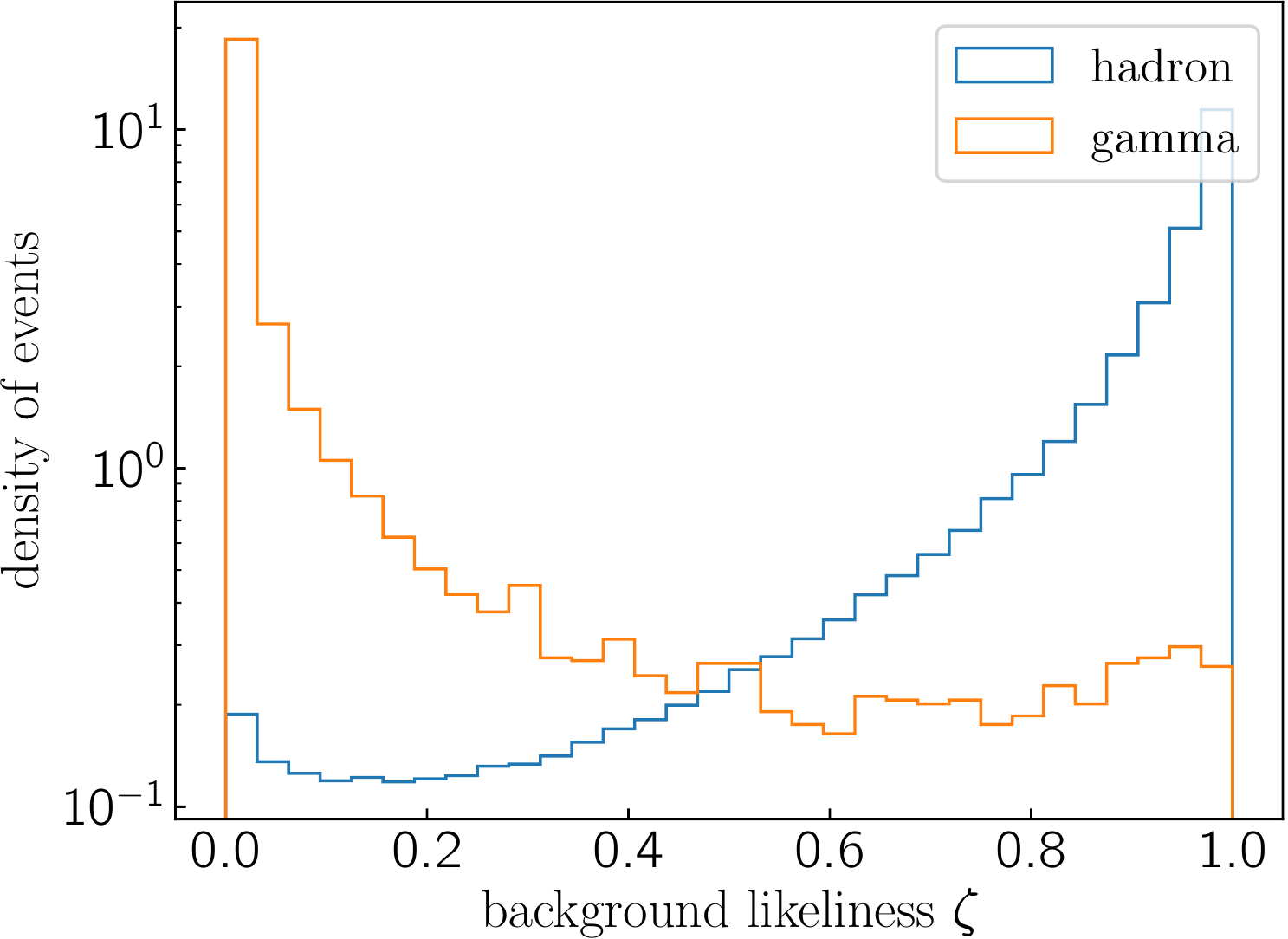}\hspace{0.5cm}
\includegraphics[height=0.31\linewidth]{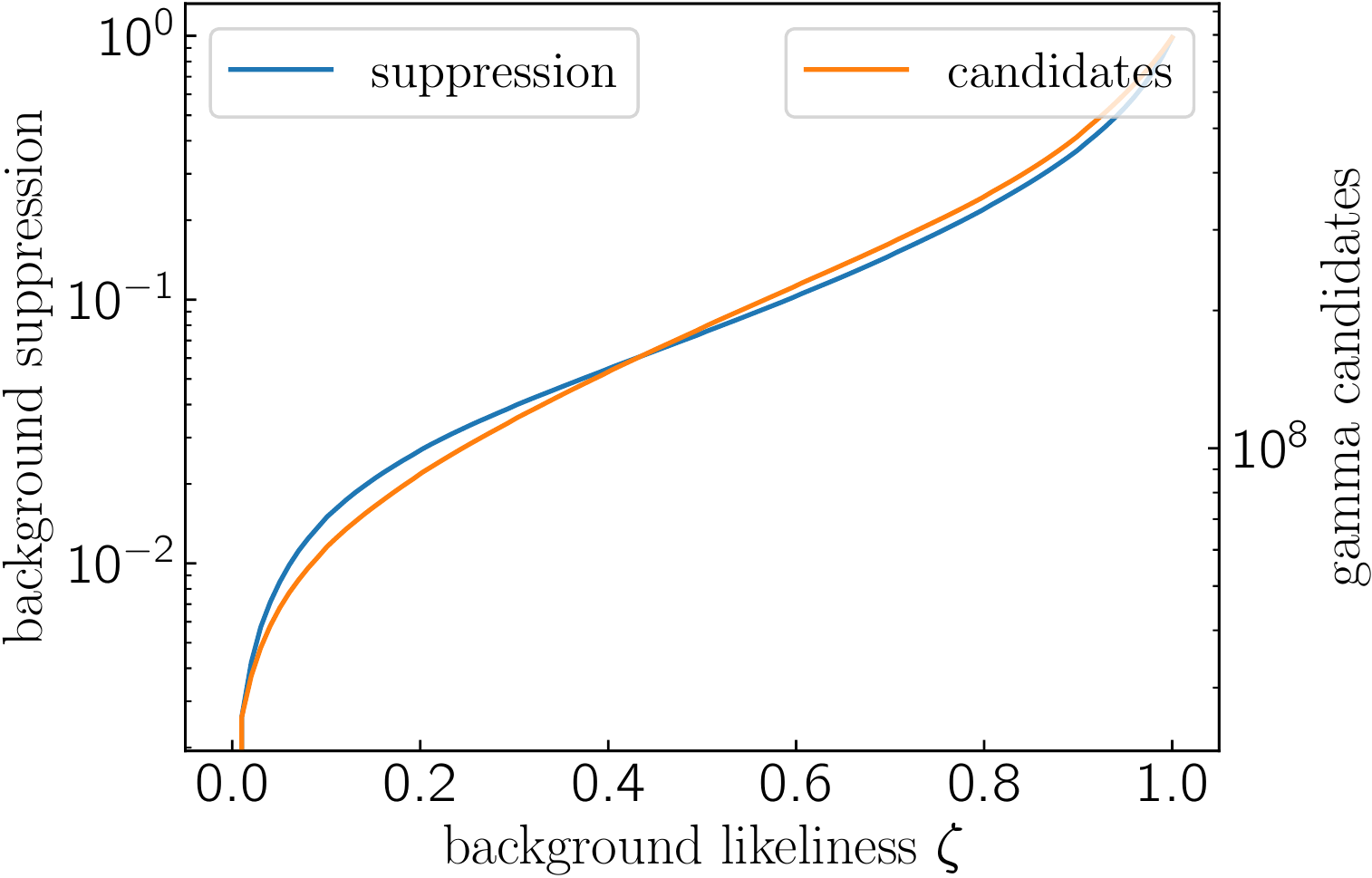}
\caption{Performance of the photon classifier developed for KASCADE data. 
\textit{Left:} Distribution of classifier output $\zeta$ for primary hadrons and photons. 
\textit{Right:} Fraction of KASCADE events classified as photons as function of $\zeta$.}
\label{fig:zeta}
\end{figure}

\section{Towards a Search for PeV Gamma-Rays}\label{sec5}

The search for high-energy $\gamma$-rays in the PeV domain is of special interest. The absorption length of PeV $\gamma$-rays via pair production in the cosmic microwave background is of the order of 10~kpc, comparable to the distance of the solar system to the Galactic Center. Only Galactic PeVatrons are visible via this channel, consistent with recent observations by HAWC~\cite{HAWC:2019tcx}, Tibet-AS$\gamma$~\cite{TibetASgamma:2021tpz} and LHAASO~\cite{LHAASO2021}.

Since the expected fraction of PeV $\gamma$-rays are a few orders of magnitude lower than the hadronic background, the binary classification approach will not provide significant detection (the classifier provides only 1:10 suppression). In order to increase the efficiency of hadron separation we use a random forest regressor returning a predicted floating point value $\zeta \in [0,1]$ which can be treated as class membership probability of reconstructed event. This allows us to choose a threshold for $\zeta$ that optimizes the signal-to-background ratio (see left panel of Fig.~\ref{fig:zeta}). The random forest consisting of 1000 trees gives us a suppression power in the range $10^2$--$10^3$. When running the classifier on real data we see that the number of $\gamma$-ray candidates corresponds to the suppression power (right panel of Fig.~\ref{fig:zeta}), but the method requires further optimization.

\section{Conclusion}\label{sec6}

We have presented the first results of a novel mass composition analysis based on archival data of the KASCADE air shower experiment acquired from 1998 to 2013 and provided by the KASCADE Cosmic ray Data Center (KCDC). 
Using modern machine learning techniques trained on data features provided by KCDC we have obtained CR spectra for five mass groups represented by \textit{H}, \textit{He}, \textit{C}, \textit{Si}, \textit{Fe}, using latest hadronic models and machine learning algorithms. This allows us to perform cross-checks with a state-of-the-art reconstruction recently published by IceCube/IceTop~\cite{IceCube:2019hmk}.
For the first time, we performed a reconstruction of large-scale anisotropy of CRs in the PeV energy domain as function of rigidity.
This intermediate success drives us to move towards search for ultra-high energy photons in KASCADE data.

The results presented in these proceedings are only the first step in our reanalysis of archival KASCADE data provided by KCDC. In future work, we plan to use deep neural networks taking single station responses as input, which are also provided by KCDC. Moreover, we plan to include KASCADE-Grande data in order to push towards higher energies.

Last but not least, we were able to outreach our activity and participate in JetBrains internship program\footnote{\url{https://internship.jetbrains.com/projects/994/}} and mathematical workshop organized by NSU\footnote{\url{https://bmm.mca.nsu.ru/project/33}}, thanks to the FAIR-ness\footnote{\url{https://www.go-fair.org/}} of KCDC data. We are preparing software and data release related for this analysis, some tutorial notebooks are already avalable in Jupyter Hub at IAP KIT\footnote{\url{https://jupyter.iap.kit.edu/}}.

\section*{Acknowledgements}
\vspace{-0.2cm}
\noindent The authors would like to express gratitude to the colleagues from KCDC team.  M.A.~acknowledges support from \textsc{Villum Fonden} under project no.~18994. The development and testing of the classifier was supported by the state contract with Institute of Thermophysics SB RAS.

\bibliographystyle{utphys_mod}
\bibliography{bibliography}

\end{document}